\documentclass[onecolumn,amsmath,amssymb]{revtex4}

\usepackage{graphicx}
\usepackage{dcolumn}
\usepackage{bm}

\begin{document}

\title{Markovian and Post-Markovian dynamics of nonequilibrium
thermal entanglement}

\author{Ilya Sinayskiy (Sinaysky)}
 \email{ilsinay@gmail.com}
\affiliation{ Quantum Research Group, School of Physics,
 University of KwaZulu-Natal, Durban, 4001, South Africa}

\author{Francesco Petruccione}
 \email{petruccione@ukzn.ac.za}
\affiliation{ Quantum Research Group, School of Physics and
National Institute for Theoretical Physics,
 University of KwaZulu-Natal, Durban, 4001, South Africa}
\date{\today}

\begin{abstract}
The dynamics of two spins coupled to bosonic baths at different
temperatures is studied. The analytical solution for the reduced
density matrix of the system in the Markovian and Post-Markovian
case with exponential memory kernel is found. The dynamics and
temperature dependence of spin-spin entanglement is analyzed.
\end{abstract}
\maketitle

\section{Introduction}

The influence of the environment plays an essential role in the
description of the realistic quantum system ~\cite{toqs}. Usually,
environment destroy entanglement in the subsystem of interest.
However in some cases it can create quantum correlations in the
system ~\cite{braun, diehl, vedral}. One of the ways to understand
the role of the parameters of the system is to study exactly
solvable models. Here, we study the dynamics of a model that was
recently introduced by L. Quiroga~\cite{Qui}. It consists of two
interacting spins in contact with two reservoirs at different
temperatures. In such a non-equilibrium case most studies are
restricted to the steady-state solutions
~\cite{Qui,prosen,key-6,medi} or to the zero temperature limit
~\cite{key-7}.

This paper is organized as follows. In Sec. II we describe the
model of a spin chain coupled to bosonic baths at different
temperatures and derive a master equation in Born-Markov
approximation. In Sec. III we present the analytical solution for
the system dynamics in the Markovian case, details of the solution
can be found in Ref.~\cite{qph}. In Sec. IV we present the
analytical solution of the Post-Markovian master equation recently
introduced by Shabani and Lidar  ~\cite{Lidar}. Finally, in Sec.
V we discuss the results and conclude.

\section{Model}

We consider a system of two interacting spins, with each spin
coupled to a separate bosonic bath. The total Hamiltonian is
given by
$$
\hat{H}=\hat{H}_{S}+\hat{H}_{B1}+\hat{H}_{B2}+\hat{H}_{SB1}+\hat{H}_{SB2},$$
where \[
\hat{H}_{S}=\frac{\epsilon_{1}}{2}\hat{\sigma}_{1}^{z}+\frac{\epsilon_{2}}{2}\hat{\sigma}_{2}^{z}+K(\hat{\sigma}_{1}^{+}\hat{\sigma}_{2}^{-}+\hat{\sigma}_{1}^{-}\hat{\sigma}_{2}^{+})\]
 is the Hamiltonian describing spin-to-spin interactions and $\hat{\sigma}_{i}^{z},\hat{\sigma}_{i}^{\pm}$
are the Pauli matrices. Note, that the units are chosen such that
$k_B=\hbar=1$. The constants $\epsilon_{1}$ and $\epsilon_{2}$
denote the energy of spins 1 and 2, respectively and $K$ denotes
the strength of the spin-spin interaction. The Hamiltonians of
the bosonic "baths" for each spin $j=1,2$ are given by
$$
\hat{H}_{Bj}=\sum_n \omega_{n,j}\hat{b}^\dag_{n,j}\hat{b}_{n,j}.
$$
 The interaction between the spin subsystem and the reservoir with creation operators $\hat{b}^\dag_{n,j}$ is described by
$$
\hat{H}_{SBj}=\hat{\sigma}_{j}^{+}\sum_{n}g_{n}^{(j)}\hat{b}_{n,j}+\hat{\sigma}_{j}^{-}\sum_{n}g_{n}^{(j)*}\hat{b}_{n,j}^{\dag}\equiv\sum_{\mu}\hat{V}_{j,\mu}\hat{f}_{j,\mu}.$$
The operators of the transitions in dynamical subsystem
$\hat{V}_{j,\mu}$ are chosen to satisfy
$[\hat{H}_{S},\hat{V}_{j,\mu}]=\omega_{j,\mu}\hat{V}_{j,\mu},$ and
the $\hat{f}_{j,\mu}$ act on the reservoir degrees of freedom.
The total system (two spins with reservoirs) is described by the
Liouville equation
\[ \frac{d}{dt}\hat{\alpha}=-i[\hat{H},\hat{\alpha}].\] We assume
that the evolution of the dynamical subsystem (coupled spins) does
not influence the state of the environment (bosonic reservoirs) so
that the density operator of the whole system $\hat{\alpha}(t)$
can be written as: \[
\hat{\alpha}(t)=\hat{\rho}(t)\hat{B}_{1}(0)\hat{B}_{2}(0),\]
where each bosonic bath is described by a canonical density matrix
$\hat{B}_{j}=e^{-\beta_{j}\hat{H}_{Bj}}/\textrm{tr}[e^{-\beta_{j}\hat{H}_{Bj}}]$
and $\hat{\rho}(t)$ denotes the reduced density matrix of the
spin subsystem.

In Born-Markov approximation the equation for the evolution of the
reduced density matrix \cite{goldman} is: \[
\frac{d\hat{\rho}}{dt}=-i[\hat{H}_{S},\hat{\rho}]+{\cal
L}_{1}(\hat{\rho})+{\cal L}_{2}(\hat{\rho})\],
 with dissipators

\begin{eqnarray*}
\mathcal{L}_{j}(\hat{\rho})\equiv\sum_{\mu,\nu}J_{\mu,\nu}^{(j)}(\omega_{j,\nu})\{[\hat{V}_{j,\mu},[\hat{V}_{j,\nu}^{\dag},\hat{\rho}]]
-(1-e^{\beta_{j}\omega_{j,\nu}})[\hat{V}_{j,\mu},\hat{V}_{j,\nu}^{\dag}\hat{\rho}]\}\end{eqnarray*}
 and where the spectral density is given by

\[
J_{\mu,\nu}^{(j)}(\omega_{j,\nu})=\int_{0}^{\infty}dse^{i\omega_{j,\nu}s}\langle
e^{-is\hat{H}_{Bj}}\hat{f}_{j,\nu}^{\dag}e^{is\hat{H}_{Bj}}\hat{f}_{j,\mu}\rangle_{j}.\]
To find a solution we go to the basis of the eigenvectors
$|\lambda_{i}\rangle$ with eigenvalues $\lambda_{i}$ of the
Hamiltonian $\hat{H}_{S},$
\[
|\lambda_{1}\rangle=|0,0\rangle,\quad\lambda_{1}=-\frac{\epsilon_1+\epsilon_2}{2},\]
 \[
|\lambda_{2}\rangle=|1,1\rangle,\quad\lambda_{2}=\frac{\epsilon_1+\epsilon_2}{2},\]
 \[
|\lambda_{3}\rangle={\rm cos}(\theta/2)|1,0\rangle+{\rm
sin}(\theta/2)|0,1\rangle,\quad\lambda_{3}=\kappa,\]
 \[
|\lambda_{4}\rangle=-{\rm sin}(\theta/2)|1,0\rangle+{\rm
cos}(\theta/2)|0,1\rangle,\quad\lambda_{4}=-\kappa,\]
 where $\kappa\equiv\sqrt{K^{2}+\frac{(\Delta\epsilon)^{2}}{4}}$
and ${\rm tan}\theta\equiv2K/(\Delta\epsilon)$. In this
representation the dissipators ${\cal L}_{i}(\hat{\rho})$ becomes
\[ {\cal
L}_{j}(\hat{\rho})=\sum_{\mu=1}^{2}J^{(j)}(-\omega_{\mu})(2\hat{V}_{j,\mu}\hat{\rho}\hat{V}_{j,\mu}^{\dag}-\{\hat{\rho},\hat{V}_{j,\mu}^{\dag}\hat{V}_{j,\mu}\}_+)\]
 \[
+J^{(j)}(\omega_{\mu})(2\hat{V}_{j,\mu}^{\dag}\hat{\rho}\hat{V}_{j,\mu}-\{\hat{\rho},\hat{V}_{j,\mu}\hat{V}_{j,\mu}^{\dag}\}_+),\]
 with transition frequencies\[
\omega_{1}=\lambda_{2}-\lambda_{3},\]
\[\omega_{2}=\lambda_{2}+\lambda_{3}\]
and transition operators \[ \hat{V}_{1,1}={\rm
cos}(\theta/2)(|\lambda_{1}\rangle\langle\lambda_{3}|+|\lambda_{4}\rangle\langle\lambda_{2}|),\]
 \[ \hat{V}_{1,2}={\rm
sin}(\theta/2)(|\lambda_{3}\rangle\langle\lambda_{2}|-|\lambda_{1}\rangle\langle\lambda_{4}|),\]
 \[\hat{V}_{2,1}={\rm
sin}(\theta/2)(|\lambda_{1}\rangle\langle\lambda_{3}|-|\lambda_{4}\rangle\langle\lambda_{2}|)
,\]
 \[\hat{V}_{2,2}={\rm
cos}(\theta/2)(|\lambda_{3}\rangle\langle\lambda_{2}|+|\lambda_{1}\rangle\langle\lambda_{4}|)
.\] In this paper we consider the bosonic bath as an infinite set
of harmonic oscillators, so the spectral density has the form
$J^{(j)}(\omega_{\mu})=\gamma_{j}(\omega_{\mu})n_{j}(\omega_{\mu})$,
where $n_{j}(\omega_{\mu})=(e^{\beta_{j}\omega_{\mu}}-1)^{-1}$ and
$J^{(j)}(-\omega_{\mu})=e^{\beta_{j}\omega_{\mu}}J^{(j)}(\omega_{\mu})$.
For simplicity we choose the coupling constant to be frequency
independent $\gamma_{1}(\omega)=\gamma_1$ and
$\gamma_{2}(\omega)=\gamma_2.$ In  the basis $|\lambda_i\rangle$
the equation for the diagonal elements of the reduced density
matrix is given by \[ \frac{d}{dt}\left(\begin{array}{c}
\rho_{11}(t)\\
\rho_{22}(t)\\
\rho_{33}(t)\\
\rho_{44}(t)\end{array}\right)=B\left(\begin{array}{c}
\rho_{11}(t)\\
\rho_{22}(t)\\
\rho_{33}(t)\\
\rho_{44}(t)\end{array}\right),\] where $B$ is a $4\times4$ matrix
with constant coefficients.
 The time-dependence for the non-diagonal elements has the following
form

$$
\rho_{i,j}(t)=e^{ts_{i,j}}\rho_{i,j}(0),$$ where $s_{i,j}$ is a
complex number. For the initial state of the system in the
computational basis $\left\{
|00\rangle,|01\rangle,|10\rangle,|11\rangle\right\} $ we
choose\begin{eqnarray*}
\lefteqn{\hat{\rho}(0)=p_{0}|00\rangle\langle00|+p_{1}|01\rangle\langle01|+p_{2}|10\rangle\langle10|}\\
 & +(1-p_{0}-p_{1}-p_{2})|11\rangle\langle11|+c_{12}|01\rangle\langle10|+c_{12}^{*}|10\rangle\langle01|.\end{eqnarray*}

\section{Exact solution in the Markovian case}

The analytical solution in the basis of eigenvectors
$|\lambda_{i}\rangle$ is given by:

\[
\rho_{ii}(t)=\frac{1}{X_{1}Y_{2}}\left(\begin{array}{cccc}
a_{11} & a_{12} & a_{13} & a_{14}\\
a_{21} & a_{22} & a_{23} & a_{24}\\
a_{31} & a_{32} & a_{33} & a_{34}\\
a_{41} & a_{42} & a_{43} &
a_{44}\end{array}\right)\left(\begin{array}{c}
\rho_{11}(0)\\
\rho_{22}(0)\\
\rho_{33}(0)\\
\rho_{44}(0)\end{array}\right),\]
 where the coefficients $a_{ij}$ are given by:
 \[a_{11}=(X_{1}^{+}+X_{1}^{-}e^{-tX_{1}})(Y_{2}^{+}+Y_{2}^{-}e^{-tY_{2}}),\]
 \[a_{12}=(1-e^{-tX_{1}})(1-e^{-tY_{2}})X_{1}^{+}Y_{2}^{+},\]
 \[a_{13}=(1-e^{-tX_{1}})X_{1}^{+}(Y_{2}^{+}+Y_{2}^{-}e^{-tY_{2}}),\]
 \[a_{14}=(X_{1}^{+}+X_{1}^{-}e^{-tX_{1}})(1-e^{-tY_{2}})Y_{2}^{+},\]
 \[a_{21}=(1-e^{-tX_{1}})(1-e^{-tY_{2}})X_{1}^{-}Y_{2}^{-},\]
 \[a_{22}=(X_{1}^{-}+X_{1}^{+}e^{-tX_{1}})(Y_{2}^{-}+Y_{2}^{+}e^{-tY_{2}}),\]
 \[a_{23}=(X_{1}^{-}+X_{1}^{+}e^{-tX_{1}})(1-e^{-tY_{2}})Y_{2}^{-},\]
 \[a_{24}=(1-e^{-tX_{1}})X_{1}^{-}(Y_{2}^{-}+Y_{2}^{+}e^{-tY_{2}}),\]
 \[a_{31}=(1-e^{-tX_{1}})X_{1}^{-}(Y_{2}^{+}+Y_{2}^{-}e^{-tY_{2}}),\]
 \[a_{32}=(X_{1}^{-}+X_{1}^{+}e^{-tX_{1}})(1-e^{-tY_{2}})Y_{2}^{+},\]
 \[a_{33}=(X_{1}^{-}+X_{1}^{+}e^{-tX_{1}})(Y_{2}^{+}+Y_{2}^{-}e^{-tY_{2}}),\]
 \[a_{34}=(1-e^{-tX_{1}})(1-e^{-tY_{2}})X_{1}^{-}Y_{2}^{+},\]
 \[a_{41}=(X_{1}^{+}+X_{1}^{-}e^{-tX_{1}})(1-e^{-tY_{2}})Y_{2}^{-},\]
 \[a_{42}=(1-e^{-tX_{1}})X_{1}^{+}(Y_{2}^{-}+Y_{2}^{+}e^{-tY_{2}}),\]
 \[a_{43}=(1-e^{-tX_{1}})(1-e^{-tY_{2}})X_{1}^{+}Y_{2}^{-},\]
 \[a_{44}=(X_{1}^{+}+X_{1}^{-}e^{-tX_{1}})(Y_{2}^{-}+Y_{2}^{+}e^{-tY_{2}}).\]
Taking into account the initial conditions, the non-vanishing
non-diagonal elements are: \[
\rho_{34}(t)=\exp\left({-i2t\lambda_{3}-\frac{t(X_{1}+Y_{2})}{2}}\right)\rho_{34}(0),\]
 \[
\rho_{43}(t)=\bar{\rho}_{34}=\exp\left({i2t\lambda_{3}-\frac{t(X_{1}+Y_{2})}{2}}\right)\rho_{43}(0).\]
 In the solution we have introduced some constants: \[
X_{i}=X_{i}^{+}+X_{i}^{-},\]
\[Y_{i}=Y_{i}^{+}+Y_{i}^{-},\]
$$ X_{i}^{\mp}=2{\rm cos}^2(\theta/2)J^{(1)}(\pm\omega_{i})+2{\rm sin}^2(\theta/2)J^{(2)}(\pm\omega_{i})$$
$$ Y_{i}^{\mp}=2{\rm sin}^2(\theta/2)J^{(1)}(\pm\omega_{i})+2{\rm cos}^2(\theta/2)J^{(2)}(\pm\omega_{i})$$
or
\begin{eqnarray*}
X_{i}^{\mp} = (J^{(1)}(\pm\omega_{i})+J^{(2)}(\pm\omega_{i}))
+\frac{\Delta\epsilon}{\sqrt{4K^2+(\Delta\epsilon)^2}}(J^{(1)}(\pm\omega_{i})-J^{(2)}(\pm\omega_{i}))
\end{eqnarray*}
\begin{eqnarray*}
Y_{i}^{\mp} =(J^{(1)}(\pm\omega_{i})+J^{(2)}(\pm\omega_{i}))
-\frac{\Delta\epsilon}{\sqrt{4K^2+(\Delta\epsilon)^2}}(J^{(1)}(\pm\omega_{i})-J^{(2)}(\pm\omega_{i})).\end{eqnarray*}
One can easily see that the only steady-state solution possible
in this system corresponds to the time moment $t=\infty$:
 \[
\lim_{t\rightarrow\infty}\rho_{ii}(t)=\frac{1}{X_{1}Y_{2}}\left(\begin{array}{c}
X_{1}^{+}Y_{2}^{+}\\
X_{1}^{-}Y_{2}^{-}\\
X_{1}^{-}Y_{2}^{+}\\
X_{1}^{+}Y_{2}^{-}\end{array}\right),\]

\[
\lim_{t\rightarrow\infty}\rho_{34}(t)=0.\]

In the regular basis $\rho_{\infty}$ is:

\[
\rho_{\infty}=\frac{1}{X_{1}Y_{2}}\times
\]

\[
\left(\begin{array}{cccc}
X_{1}^{-}Y_{2}^{-} &   0    &    0   &   0   \\
  0    & {\rm c}^2 X_{1}^{-}Y_{2}^{+}+{\rm s}^2X_{1}^{+}Y_{2}^{-} & {\rm s}(X_{1}^{-}Y_{2}^{+}-X_{1}^{+}Y_{2}^{-}) &   0   \\
  0    & {\rm s}(X_{1}^{-}Y_{2}^{+}-X_{1}^{+}Y_{2}^{-}) & {\rm s}^2X_{1}^{-}Y_{2}^{+}+{\rm c}^2X_{1}^{+}Y_{2}^{-} &   0   \\
  0    &   0    &    0   & X_{1}^{+}Y_{2}^{+}\end{array}\right),
\]
where $c=\cos\left(\theta/2\right)$ and
$s=\sin\left(\theta/2\right)$.

In order to quantify the entanglement between the spins we
consider the concurrence~\cite{woot}. In the steady-state
$(t\rightarrow\infty)$ it is given by\begin{eqnarray*}
C_{\infty}=\frac{2}{X_1Y_2}\textrm{Max}\left(0,\frac{{\rm
sin}\theta}{2}|X_1^+Y_2^--X_1^-Y_2^+|-\sqrt{X_1^-X_1^+Y_2^-Y_2^+}\right)
.\end{eqnarray*}

\section{Exact solution in the Post-Markovian case}

It is a well known fact that positivity is guaranteed only in the
case of the Markovian dynamics and in general even in the
Born-approximation one can find that evolution in no longer
positive. Recently Shabani and Lidar \cite{Lidar,lidar2}
suggested and studied an equation which describes positive and
non-Markovian dynamics of the reduced system, so called
Post-Markovian dynamics

\[
\frac{d\rho}{dt}=\mathcal{L}\int_0^tdt'k(t')\exp{(t'\mathcal{L})}\rho(t-t'),\]
or
\[
\frac{d\rho}{dt}=\mathcal{L}k(t)\exp{(t\mathcal{L})}\ast\rho(t).\]
Note, that the above dynamics contains a phenomenological memory
kernel $k(t)$.

Solution of the post-Markovian equation can be constructed with
the help of the Laplace transform
\[
s\rho(s)-\rho(0)=[k(s)\ast\frac{\mathcal{L}}{s-\mathcal{L}}]\rho(s).
\]
Then, the eigenvector-problem for the Lindbladian
\[
\mathcal{L}\rho=\lambda\rho
\]
can be solved and one gets the solution
\[
\rho(t)=\sum_i Tr[L_i\rho(t)]R_i=\sum_i \mu_i(t)R_i,
\]
where
\[
\mu_i(t)=\textrm{Lap}^{-1}[\frac{1}{s-\lambda_i
k(s-\lambda_i)}]\mu_i(0)=\xi_i(t)\mu_i(0).
\]
Particularly, in the case considered in this article the
post-Markovian equation takes the following form:
\[
\frac{d\rho}{dt}=-i[H_s,\rho]+(\mathcal{L}_1+\mathcal{L}_2)\int_0^tdt'k(t')\exp{(t'(\mathcal{L}_1+\mathcal{L}_2))}\rho(t-t').\]
In order to solve the eigenvector problem we find the Jordan
decomposition for the Lindbladian
\[
\mathcal{L}=\mathcal{L}_1+\mathcal{L}_2=S J S^{-1},
\]
where
\[
S=\left(\begin{array}{cccc}
Y_2^+/Y_2^- & Y_2^+/Y_2- & -1 & -1 \\
X_1^-/X_1^+ & -1  & X_1^-/X_1^+ & -1\\
X_1^-Y_2^+/X_1^+Y_2^- & -Y_2^+/Y_2^- & -X_1^-/X_1^+ & 1\\
1 & 1 & 1 & 1\end{array}\right)
\]
and
\[
J=\textrm{diag}(0,-X_1,-Y_2,-X_1-Y_2).
\]
In this paper we consider the exponential memory kernel \cite{FR}
\[
k(t)=\gamma_0 e^{-\gamma_0t}, \] which implies that
\[
\xi(\lambda_i,t)=\frac{\gamma_0e^{\lambda_it}+\lambda_ie^{-\gamma_0t}}{\gamma_0+\lambda_i}
\]

The analytical solution for the diagonal elements is
\[
\rho_{ii}(t)=S\times\textrm{diag}(1,\xi(J_{22},t),\xi(J_{33},t),\xi(J_{44},t))\times
S^{-1}\left(\begin{array}{c}
\rho_{11}(0)\\
\rho_{22}(0)\\
\rho_{33}(0)\\
\rho_{44}(0)\end{array}\right).\] Taking into account that for
the non-diagonal elements the Lindbladian has Jordan form the
dynamics of the corresponding elements are given by
$\xi(\lambda_{non-diag},t)$ with corresponding value.

\begin{figure}
\includegraphics[scale=0.7]{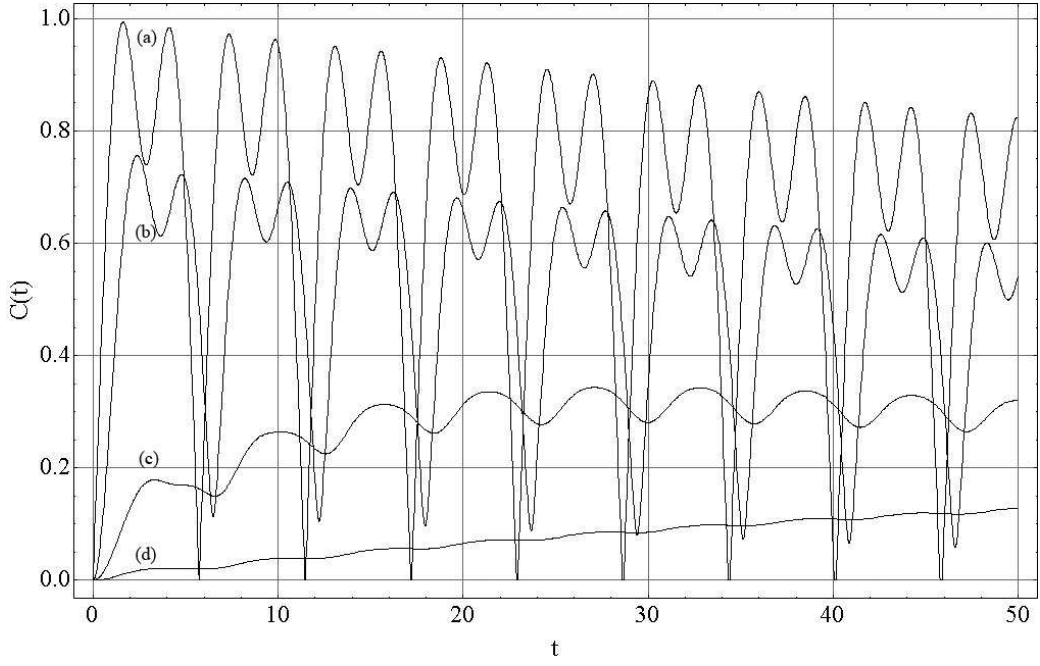}
\caption{Dynamics of the concurrence $C(t)$ for the initial
reduced density matrix $\hat{\rho}_0=|1,0\rangle\langle1,0|$. The
parameters of the model are chosen to be
$\gamma_1=\gamma_2=0.001$, $\epsilon_1=2$, $\epsilon_2=1.1$,
$K=1$, $T_1=0.2$, $T_2=0.5$ for different coefficients $\gamma_0$
in the memory kernel $k(t)$: curve $(a)$ corresponds to the
Markovian case $k(t)=\delta(t)$; curves (b)-(d) post-Markovian
cases; curve (b) $\gamma_0=1$; curve (c) $\gamma_0=0.1$; curve
(d) $\gamma_0=0.01$.}
\end{figure}

\begin{figure}
\includegraphics[scale=0.7]{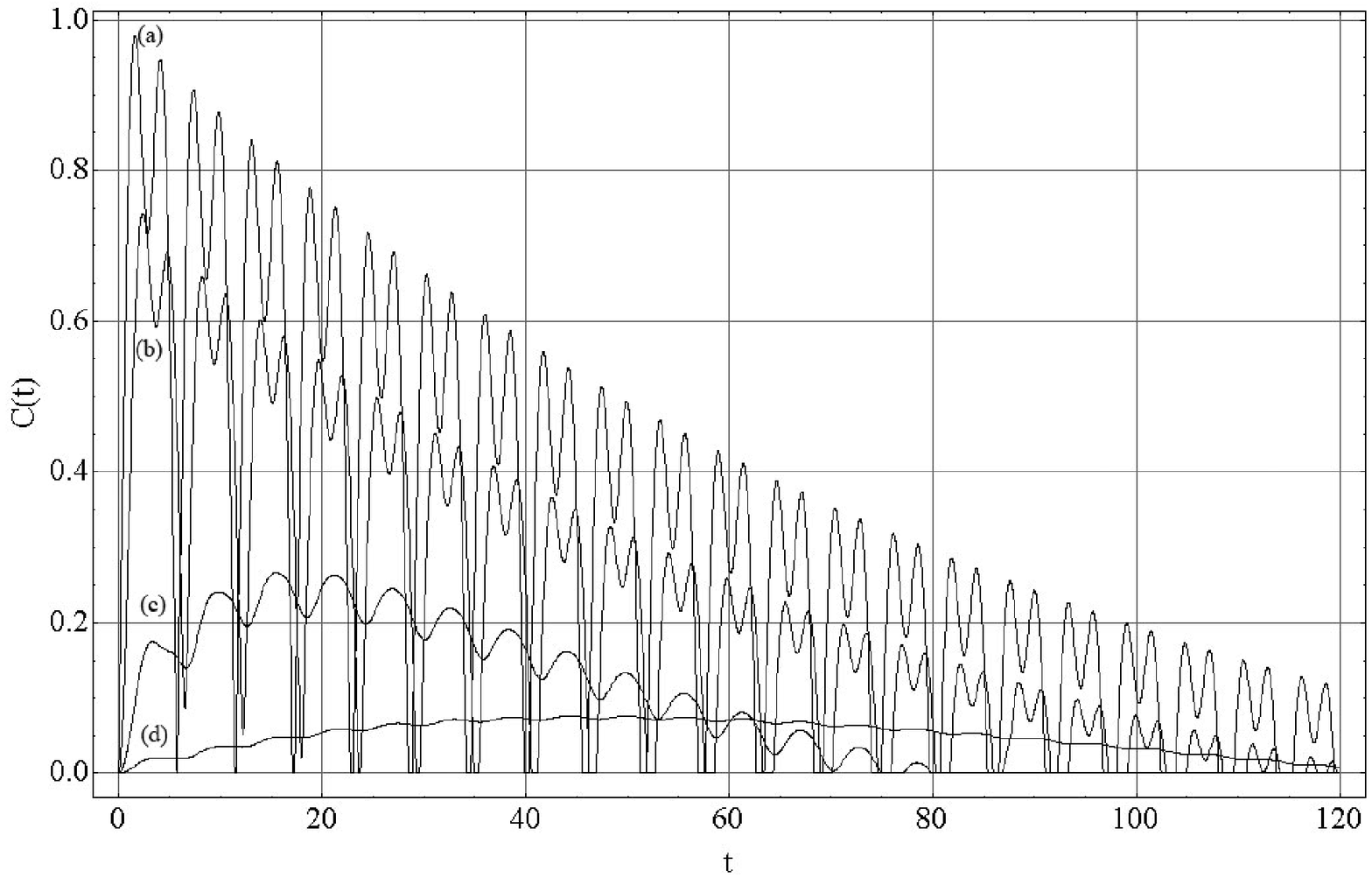}
\caption{Dynamics of the concurrence $C(t)$ for the initial
reduced density matrix $\hat{\rho}_0=|1,0\rangle\langle1,0|$. The
parameters of the model are chosen to be
$\gamma_1=\gamma_2=0.001$, $\epsilon_1=2$, $\epsilon_2=1.1$,
$K=1$, $T_1=1.2$, $T_2=1.5$ for different coefficients $\gamma_0$
in the memory kernel $k(t)$: curve $(a)$ corresponds to Markovian
case $k(t)=\delta(t)$; curves (b)-(d) post-Markovian cases; curve
(b) $\gamma_0=1$; curve (c) $\gamma_0=0.1$; curve (d)
$\gamma_0=0.01$.}
\end{figure}

\begin{figure}
\includegraphics[scale=0.7]{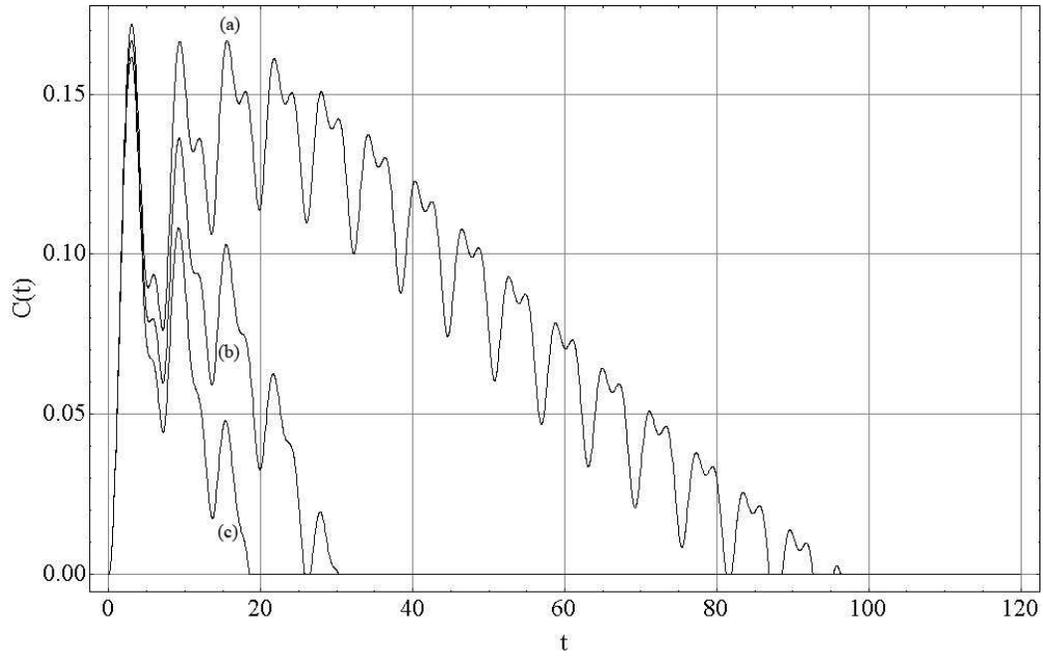}
\caption{Dynamics of the concurrence $C(t)$ for the initial
reduced density matrix $\hat{\rho}_0=|1,0\rangle\langle1,0|$. The
parameters of the model are chosen to be
$\gamma_1=\gamma_2=0.001$, $\epsilon_1=1.5$, $\epsilon_2=1.1$,
$K=1$, $\gamma_0=0.1$  for different temperatures of the "baths":
curve $(a)$ corresponds to $T_1=0.2$, $T_2=0.5$; (b) $T_1=1.2$,
$T_2=1.5$; (c) $T_1=2.2$, $T_2=2.5$.}
\end{figure}

\begin{figure}
\includegraphics[scale=0.7]{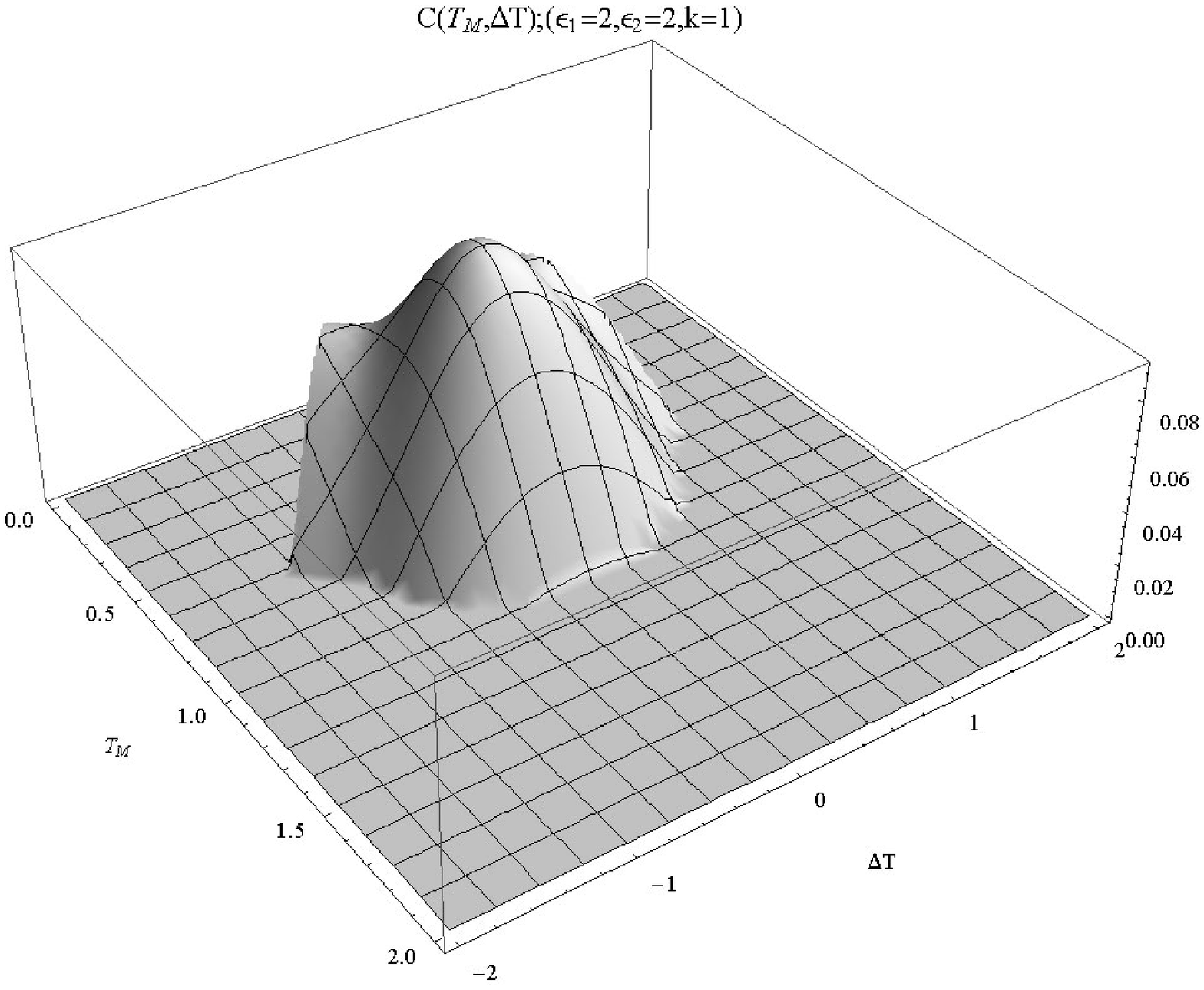}
\caption{Steady-state concurrence $C(T_M,\Delta T)$ as a function
of the mean bath temperature $T_M=(T_1+T_2)/2$ and temperature
difference $\Delta T=T_1-T_2$ in the symmetric case
$\epsilon_1=\epsilon_2=2$ with $K=1$.}
\end{figure}

\begin{figure}
\includegraphics[scale=0.7]{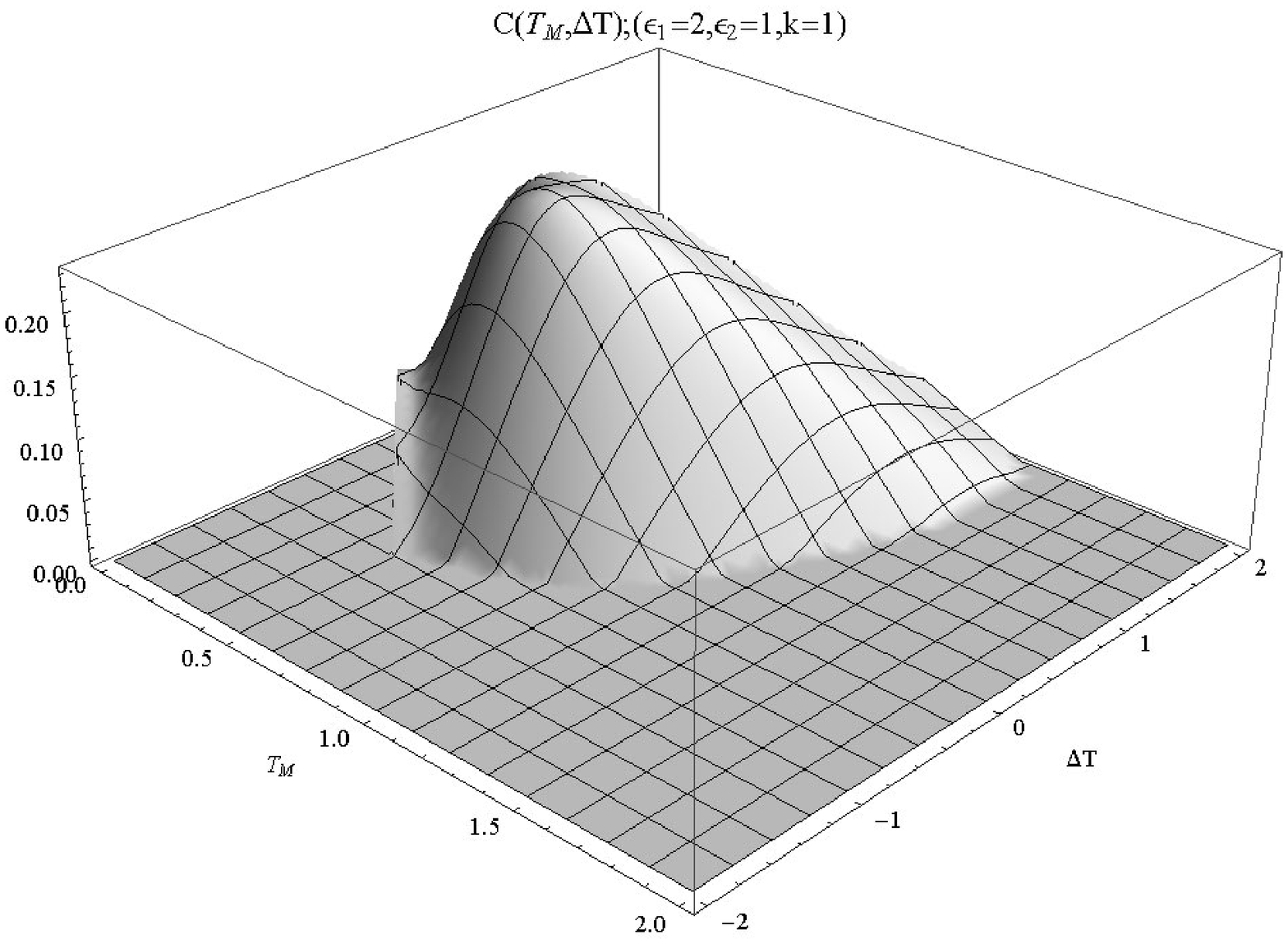}
\caption{Steady-state concurrence $C(T_M,\Delta T)$ as a function
of the mean bath temperature $T_M=(T_1+T_2)/2$ and the
temperature difference $\Delta T=T_1-T_2$ in the case
$\epsilon_1=2$, $\epsilon_2=1$, $K=1$.}
\end{figure}

\section{Results and Discussion}

The dynamics of entanglement is analyzed in Figures 1-3. In
Figures 1 and 2 the dynamics of the concurrence between the two
qubits is shown for different coefficients $\gamma_0$ in the
memory kernel $k(t)$ (for the Markovian case $\gamma_0=\infty$).
One can see that with decreasing $\gamma_0$ memory effects play a
more essential role in the system dynamics and practically
suppress the oscillations in the concurrence dynamics due to the
Hamiltonian dynamics of the system. In Figure 3 one can see that
increasing the temperature of the baths destroys quantum
correlations in the system. From Figure 3 and curve (c) on Figure
2 one can see the phenomenon of "sudden death" of entanglement
Refs.~\cite{SDE1,SDE2}. The steady-state concurrence is analyzed
in Figures 4 and 5. The detailed analysis of steady state
concurrence for this model is given in Ref.~\cite{qph}. In
Figures 4 and 5 we plot the steady-state concurrence for the
symmetric $(\Delta \epsilon=0)$ and non-symmetric $(\Delta
\epsilon\neq0)$ cases as a function the mean temperature
($T_M=(T_1+T_2)/2$) and the temperature difference ($\Delta
T=T_1-T_2$) of the baths. One can see that in the symmetric case
(Fig. 4) the maximal value of the entanglement corresponds to the
thermodynamically equilibrium case $(T_1=T_2)$ and in the
non-symmetric case (Fig. 5) the maximum of the quantum
correlations reaches in the thermodynamically non-equilibrium
case.

In conclusion, we have found an analytical solution for a simple
spin system coupled to bosonic baths at different temperatures in
Markovian and Post-Markovian cases. We studied the influence of
memory effect on the dynamics of entanglement.

\end{document}